\documentclass[fleqn,usenatbib]{mnras}
\usepackage{newtxtext,newtxmath}

\usepackage[T1]{fontenc}
\usepackage{times}

\usepackage{graphicx}	
\usepackage{amsmath}	
\usepackage{amssymb}    

\usepackage{import}     
\usepackage{mathtools}  
\usepackage{microtype}  
\usepackage{ragged2e}
\usepackage{tikz}       
\usepackage{xparse}     

\hypersetup{unicode}

\newcommand{\Rv}{{\ensuremath\text{R}_{200}}}
\newcommand{\Mv}{{\ensuremath\text{M}_{200}}}
\newcommand{\Msun}{{\ensuremath\text{M}_\odot}}
\newcommand{\specline}[2]{\ensuremath{\mathchoice{{\text{#1}}\,{#2}}
                                                 {{\text{#1}}\,{#2}}
                                                 {{\text{#1}}{#2}}
                                                 {{\text{#1}}{#2}}}}
\newcommand{\Ha}{{\specline{H}{\alpha}}}
\newcommand{\Lya}{{\specline{Ly}{\alpha}}}
\newcommand{\Hmol}{{\ensuremath\text{H}_2}}
\newcommand{\spec}[2]{{\text{#1}\,\textsc{#2}}}
\newcommand{\HI}{\spec{H}{i}}
\newcommand{\HII}{\spec{H}{ii}}
\newcommand{\HeI}{\spec{He}{i}}
\newcommand{\HeII}{\spec{He}{ii}}

\newcommand{\sbunit}{\text{erg}\,\text{s}^{-1}\text{cm}^{-2}\text{arcsec}^{-2}}

\NewDocumentCommand \df {m O{d} o}{\text{#2}\IfValueT{#3}{^{#3}}#1}
\NewDocumentCommand \dff {G{} m o}{\frac{\df{#1}[#3]}{\df{#2}[#3]}}

\DeclareMathOperator{\Forall}{\forall}

\makeatletter
\newcommand*\rel@kern[1]{\kern#1\dimexpr\macc@kerna}
\newcommand*\widebar[1]{%
  \begingroup
  \def\mathaccent##1##2{%
    \rel@kern{0.8}%
    \overline{\rel@kern{-0.8}\macc@nucleus\rel@kern{0.2}}%
    \rel@kern{-0.2}%
  }%
  \macc@depth\@ne
  \let\math@bgroup\@empty \let\math@egroup\macc@set@skewchar
  \mathsurround\z@ \frozen@everymath{\mathgroup\macc@group\relax}%
  \macc@set@skewchar\relax
  \let\mathaccentV\macc@nested@a
  \macc@nested@a\relax111{#1}%
  \endgroup
}
\makeatother

\defcitealias{benitez-llambayPropertiesDarkLCDM2017}{BL17}
\defcitealias{madauCosmicReionizationPlanck2015}{MH15}
\defcitealias{sternbergAtomicHydrogenGas2002}{S02}
\defcitealias{cookePrimordialAbundanceDeuterium2016}{CP16}

\title[Fluorescent rings in star-free dark matter haloes]{Fluorescent rings in star-free dark matter haloes}

\author[C. Sykes \emph{et. al.}]{%
Calvin Sykes$^{1,2}$\thanks{E-mail: calvin.v.sykes@durham.ac.uk},
Michele Fumagalli$^{1,2}$,
Ryan Cooke$^{2}$,
\newauthor
Tom Theuns$^{1}$,
Alejandro Ben\'{i}tez--Llambay$^{1}$
\\
$^{1}$Institute for Computational Cosmology, Durham University, Durham DH1 3LE, UK\\
$^{2}$Centre for Extragalactic Astronomy, Durham University, Durham DH1 3LE, UK\\
}

\date{Accepted XXX. Received YYY; in original form ZZZ}

\pubyear{2019}

\begin{document}
\label{firstpage}
\pagerange{\pageref{firstpage}--\pageref{lastpage}}
\maketitle

\begin{abstract}
Photoheating of the gas in low-mass dark matter (DM) haloes prevents baryons from cooling, leaving the haloes free of stars.
Gas in these `dark' haloes remains exposed to the ultraviolet background (UVB), and so is expected to emit via fluorescent recombination lines.
We present a set of radiative transfer simulations, which model dark haloes as spherical gas clouds in hydrostatic equilibrium with a DM halo potential, and in thermal equilibrium with the UVB at redshift $z=0$.
We use these simulations to predict surface brightnesses in $\Ha$, which we show to have a characteristic ring-shaped morphology for haloes in a narrow mass range between ${\simeq} 10^{9.5}$ and $10^{9.6}\,\Msun$.
We explore how this emission depends on physical parameters such as the DM density profile and the UVB spectrum.
We predict the abundance of fluorescent haloes on the sky, and discuss possible strategies for their detection.
We demonstrate how detailed observations of fluorescent rings can be used to infer the properties of the haloes which host them, such as their density profiles and the mass-concentration relation, as well as to directly measure the UVB amplitude.
\end{abstract}




\section{Introduction}
\label{sec:intro}
A fundamental prediction of the cold dark matter (CDM) paradigm is the existence of a large number of dwarf haloes (those with mass ${\lesssim} 10^{10}\,\Msun$) resulting from the hierarchical nature of structure formation.
To achieve consistency with observational constraints, it is necessary that only a small fraction of these low-mass haloes host luminous galaxies \citep[e.g.][]{klypinWhereAreMissing1999, mooreDarkMatterSubstructure1999}.
This requirement can be fulfilled by the effects of baryonic feedback processes, as demonstrated by hydrodynamical simulations \citep[e.g.][]{babulDwarfEllipticalGalaxies1992,efstathiouSuppressingFormationDwarf1992}.
Specifically, cosmic reionisation heats baryonic matter to ${\sim} 10^4\;\text{K}$
(see \citealt{meiksinPhysicsIntergalacticMedium2009} for a review), inhibiting baryons from condensing into low-mass haloes to form stars.
Furthermore, gas which does accrete can later be expelled by external ram-pressure stripping as the halo moves through the intergalactic medium~\citep{benitez-llambayDwarfGalaxiesCosmic2013}.

Detection of these star-free `dark' haloes would therefore provide convincing evidence in support of CDM, and a number of methods for discovering them have been proposed.
The presence of dark haloes may be inferred gravitationally, either via their dynamical influence on other objects~\citep{erkalPropertiesDarkSubhaloes2015,feldmannDetectingDarkMatter2015}, or via gravitational lensing \citep{vegettiDetectionDarkSubstructure2010,hezavehDetectionLensingSubstructure2016}.
Alternatively, one may exploit the fact that the haloes do contain a reservoir of almost pristine gas originating from the limited accretion that occurred before reionisation.
Observing this gas either in absorption against a luminous background source \citep{reesLymanAbsorptionLines1986}, or directly in emission, could allow the presence of the dark haloes to be inferred.

Previously, \citet[][hereafter \citetalias{sternbergAtomicHydrogenGas2002}]{sternbergAtomicHydrogenGas2002} considered the emission properties of the gas bound to dark haloes, in the context of comparing them with observed high-velocity clouds (HVCs).
This work concluded that for HVCs to be dark halo candidates, they must be `circumgalactic' objects located relatively close to the Milky Way, and pressure-confined by a hot galactic corona.
More recently, \citet[][hereafter \citetalias{benitez-llambayPropertiesDarkLCDM2017}]{benitez-llambayPropertiesDarkLCDM2017} used the \textsc{Apostle} suite of Local Group hydrodynamical simulations \citep{sawalaAPOSTLESimulationsSolutions2016} to model the state of gas within dark haloes.
This work identified two populations: `COSWEBs', haloes which are found to lose almost all their residual gas after reionisation due to ram pressure stripping; and `RELHICs', generally more isolated systems which are not subject to this additional effect, and thus have relatively constant baryonic content since reionisation.\footnote{These acronyms stand for "COSmic WEB Stripped systems" and "Reionisation-Limited $\HI$ Clouds" respectively.}
RELHICs with sufficiently high halo masses were found to develop an approximately kiloparsec-sized neutral core, with the surrounding gas remaining ionised by the diffuse ultraviolet background (UVB).

Whereas \citetalias{benitez-llambayPropertiesDarkLCDM2017} considered the prospect of detecting $\HI$ 21\,cm emission from these objects, in this work we follow \citetalias{sternbergAtomicHydrogenGas2002} by investigating a second potential source of emission: the fluorescent $\Ha$ line, in order to reassess its expected properties in light of the `extragalactic' environment for RELHICs favoured by \textsc{Apostle}.
Fluorescent emission occurs when atoms ionised by the UVB recombine to produce neutral atoms in excited states, which subsequently undergo radiative cascades to return to the ground state.
Although the $n=2 \rightarrow 1$ $\Lya$ transition is intrinsically brighter, we consider the $\Ha$ transition ($n=3 \rightarrow 2$) due to its non-resonant nature and rest-frame optical wavelength, which are especially advantageous for the ground detection of low-redshift sources.

In this work, we perform photoionisation simulations of the gas within dark haloes in the local Universe.
We calculate radial profiles of the $\Ha$ surface brightness, and investigate its dependence on physical properties of the haloes and the UVB.
These calculations use an adapted version of the code described by \citet[][hereafter \citetalias{cookePrimordialAbundanceDeuterium2016}]{cookePrimordialAbundanceDeuterium2016}, which follows the procedure outlined in \citetalias{sternbergAtomicHydrogenGas2002}.
These calculations are qualitatively similar to those performed by the \textsc{cloudy} spectral synthesis code \citep{ferland2017ReleaseCloudy2017}, but with two important additional features.
Firstly, we model gas in hydrostatic equilibrium with a dark matter halo to obtain realistic gas density profiles.
Secondly, we perform the calculations in (projected) spherical coordinates, in order to capture the radial dependence of optical depths within the cloud.

This paper is organised as follows: in Section~\ref{sec:meth}, we describe our numerical method.
We use this method in Section~\ref{sec:results} to compute grids of models with a variety of input parameters to determine the resulting radial surface brightness profiles.
We then combine our results with the \textsc{apostle} simulations in Section~\ref{sec:abund} to predict the abundance of $\Ha$-fluorescent haloes on the sky, and the prospects for their detection.
Lastly, in Section~\ref{sec:discuss} we present a discussion of our results and their significance for constraining the structure and environment of a fluorescent halo, based on its observed $\Ha$ emission.
Throughout, we assume a set of cosmological parameters ($H_0=67.3\;\text{km}\,\text{s}^{-1}\,\text{Mpc}^{-1}$, $\Omega_\Lambda=0.685$, $\Omega_{\text{M}}=0.315$, $\Omega_{\text{B}}=0.0491$) consistent with \emph{Planck} measurements \citep{planckcollaborationPlanck2013Results2014}.

\section{Simulation design}
\label{sec:meth}

We use a modified version of the spherically-symmetric ionisation balance code described in \citetalias{cookePrimordialAbundanceDeuterium2016}.
A full description of this code is given therein; here we will provide a brief summary and describe the modifications we have made.\footnote{\RaggedRight The code is made available here: \url{https://github.com/calvin-sykes/spherical_cloudy}}

\subsection{Code outline}
\label{ssec:code}

We consider clouds of primordial gas, consisting only of hydrogen and helium with abundance ratio by number $n_{\text{He}}/n_{\text{H}}=0.083$ (corresponding to a primordial helium mass fraction $Y_{\text{P}} = 0.24$).
This gas is embedded within a spherically-symmetric Navarro-Frenk-White (NFW; \citealp{navarroStructureColdDark1996}) dark matter halo, with radial density profile given by
\begin{equation}
\rho(r)=\frac{\rho_{s}}{x(1+x)^2};\qquad x\equiv \frac{r}{r_s},
\label{eq:nfw_profile}
\end{equation}
where $r_s$ and $\rho_s$ are characteristic length and density scales of the halo.
The gas is assumed to be in hydrostatic equilibrium with a potential $\varphi(r)$, such that $\df{P}(r)=-\rho_{\text{g}}(r)\df{\varphi}(r)$, where $P(r)$ and $\rho_{\text{g}}(r)$ are the gas pressure and density profiles respectively.
For simplicity, we neglect the self-gravity of the gas such that $\varphi(r)$ is defined solely by the dark matter halo, which is a reasonable assumption for the DM-dominated systems we consider.

Haloes are constructed by choosing a virial mass $\Mv$, defined as the mass contained in a sphere with average density $200\rho_{\rm{crit}}$, where $\rho_{\rm{crit}}$ is the critical density of the Universe. The radius of this sphere is the virial radius $\Rv$, such that:
\begin{equation}
\Mv{}=\frac{4\pi}{3}\Rv{}^3\times 200\rho_{\rm{crit}}.
\end{equation}
NFW haloes may be parameterised by a concentration parameter $c_{200}\equiv\Rv/r_s$.
In our fiducial models, we determine values of $c_{200}$ as a function of $\Mv$ using the mass-concentration relation of \citet{ludlowMassConcentrationRedshift2016}.
We investigate alternative mass-concentration relations and density profiles in Section~\ref{ssec:dmprof} in order to determine the dependence of our results on these assumptions.
A total gas mass $\text{M}_{\text{g}}$ is then assigned to the halo using the analytic model employed by \citetalias{benitez-llambayPropertiesDarkLCDM2017}, which determines the gas mass required for hydrostatic equilibrium as a function of $\Mv$.

Given the assumed absence of local star formation, the gas is illuminated solely by photons from the diffuse UVB.
We adopt the \citet[][hereafter MH15]{madauCosmicReionizationPlanck2015} UVB in our fiducial model, although we explore the effect of varying the shape of this spectrum in Section~\ref{ssec:alpha_uv}.
Under the assumption of spherical symmetry, the UVB irradiates the gas isotropically.
However, the gas column density has an angular dependence which varies with position, meaning that the local background radiation intensity, and hence the local photoionisation rate, is a function of both depth within the cloud and direction of the incident radiation.

In addition to photoionisation, we consider primary and secondary collisional ionisation, ionisation resulting from helium recombination radiation, and charge transfer ionisation.
We include radiative, dielectronic and charge transfer recombinations.
Ionisation equilibrium for each species is enforced by equating the appropriate ionisation and recombination processes in each case.
The rate coefficients assumed for these reactions are unmodified from those detailed in \citetalias{cookePrimordialAbundanceDeuterium2016}, with the exception of the rates for collisional ionisation, which we take from the \textsc{Chianti} atomic database \citep{dereCHIANTIAtomicDatabase1997,zannaCHIANTIAtomicDatabase2015} in order to enable calculation of the rates at temperatures ${\sim} 10^3\;\text{K}$, which may be reached in the neutral core (the lower limit in the previously-used \citet{dereIonizationRateCoefficients2007} rates was instead ${\simeq} 10^4\;\text{K}$).

The version of the code described in \citetalias{cookePrimordialAbundanceDeuterium2016} additionally assumed strict thermodynamic equilibrium, such that the temperature at each radial coordinate is that for which the computed heating and cooling rates are equal.
The heating rate includes both photoionisation heating and secondary heating by primary photoelectrons, while the cooling rate incorporates contributions from collisional excitation/ionisation cooling, recombination cooling, Brehmsstrahlung cooling and Compton cooling/heating.
We take the formulae used to determine each of these contributions from  \citet{cenHydrodynamicApproachCosmology1992}.
At low gas densities ($n_{\rm{H}} \lesssim 10^{-4.8}\,\rm{cm}^{-3}$), cooling becomes inefficient and the gas temperature will instead be that which results in a heating timescale equal to the Hubble time.
We incorporate this additional regime by switching from calculating the equilibrium temperature to using the tabulated temperature-density relations given in \citet{wiersmaEffectPhotoionizationCooling2009} when the gas density falls below this transition point.

The form of the calculations follows that described by \citetalias{sternbergAtomicHydrogenGas2002}: assuming the gas is initially isothermal and fully-ionised, we compute the gas pressure profile required to maintain hydrostatic equilibrium.
A boundary condition is required to determine the pressure profile; in common with \citetalias{benitez-llambayPropertiesDarkLCDM2017}, we use the assumption that at sufficient distance from the centre of the halo (${\simeq} 100\,\Rv$), the density should approach the cosmic mean baryon (number) density $\bar{n} \simeq 10^{-6.7} \rm{cm}^{-3}$ at $z=0$.

From the computed pressure profile, we determine the radial density distribution, and hence the intensity of the radiation field within the cloud.
We then solve for ionisation equilibrium to calculate the neutral fractions of each species as a function of radius.
The temperature profile is then derived as described above, and used to recompute the pressure profile.
This process is iterated until a convergence criterion is met, requiring that the relative change in the computed neutral fractions between successive iterations is no more than $0.1\%$ at any radial coordinate.

Once a converged ionisation structure has been found, the $\Ha$ surface brightness $\Sigma_{\Ha}$ may be computed.
At a radial position within the halo $r$, $\Ha$ photons are produced at a rate given by the volume emissivity $\varepsilon_{\Ha} = n_{\HII}(r) n_e(r) \alpha_{\Ha}(r)$, where $n_{\HII}$ and $n_e$ are the densities of ionised hydrogen and electrons respectively.
The effective recombination coefficient $\alpha_{\Ha}$ expresses the rate per unit electron and ion densities at which the $n=3 \rightarrow 2$ transition occurs.
We obtain values for this coefficient, as a function of temperature, from \citet{osterbrockAstrophysicsGaseousNebulae2006}.

For a line of sight intersecting the halo with impact parameter $b$, the projected surface brightness is found as the line integral of the emissivity along the line of sight
\begin{equation}
\Sigma_{\Ha}(b)=\frac{h \nu_{\Ha}}{2\pi} \int_b^{\mathrlap{\Rv}}{\frac{r}{\sqrt{r^2 - b^2}} n_{\HII}(r) n_e(r) \alpha_{\Ha}(r)}\,\df{r},
\end{equation}
where $h$ is Planck's constant and $\nu_{\Ha}=4.57\times 10^{14}\,\text{Hz}$ is the frequency of an $\Ha$ photon.

\subsection{Tests of numerical accuracy}
\label{ssec:tests}

To validate our code, we model a series of plane-parallel gas slabs, varying the volume and total column densities of the slab.
For each slab, we initially specify a constant temperature $T_{\text{gas}}=10^4\,\text{K}$.
In this case the ionisation structure of the slab may be predicted analytically; in particular, the surface brightness is expected to approach a constant value, in a similar manner to the result for $\Lya$ fluorescence reported by \citeauthor{gouldImagingForestLyman1996} (\citeyear{gouldImagingForestLyman1996}; see Appendix~\ref{app:ppcase}).
The use of a plane-parallel geometry also enables us to compare our results with those obtained from the spectral synthesis code \textsc{cloudy} \citep{ferland2017ReleaseCloudy2017}.

\begin{figure}
\subimport{figs/}{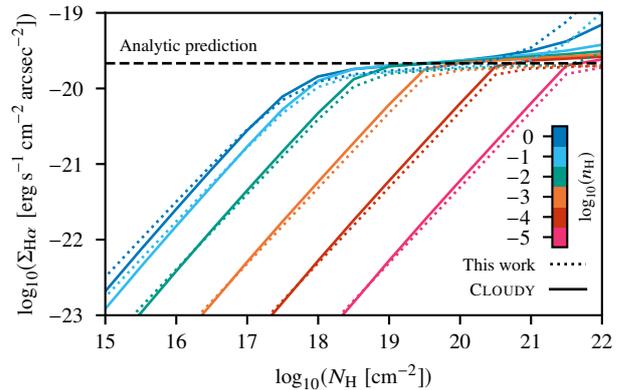}
\caption{$\Ha$ surface brightness as a function of total H column density for isothermal, H-only plane parallel models. Dotted curves are obtained from the code we describe here, solid ones from \textsc{cloudy}; each same-coloured pair corresponds to a particular constant gas density $n_{\text{H}}$. The horizontal dashed line indicates the analytic result for $\Ha$ surface brightness in the photoionised, optically-thick limit.}
\label{fig:sb_noHe_GW}
\end{figure}

Fig.~\ref{fig:sb_noHe_GW} demonstrates that in the high-column density limit, our code reproduces the analytic prediction well, with the only deviation occurring at high volume densities where collisional ionisation becomes significant.
Note that our code makes a number of simplifications as compared to \textsc{cloudy}: we assume case B recombination, rather than self-consistently computing the diffuse radiation field produced by recombinations within the cloud.
Also, our treatment of collisional processes is less sophisticated than the one in \textsc{cloudy}.
We further explore the effects of these approximations in Appendix~\ref{app:moretests}.

\begin{figure}
\subimport{figs/}{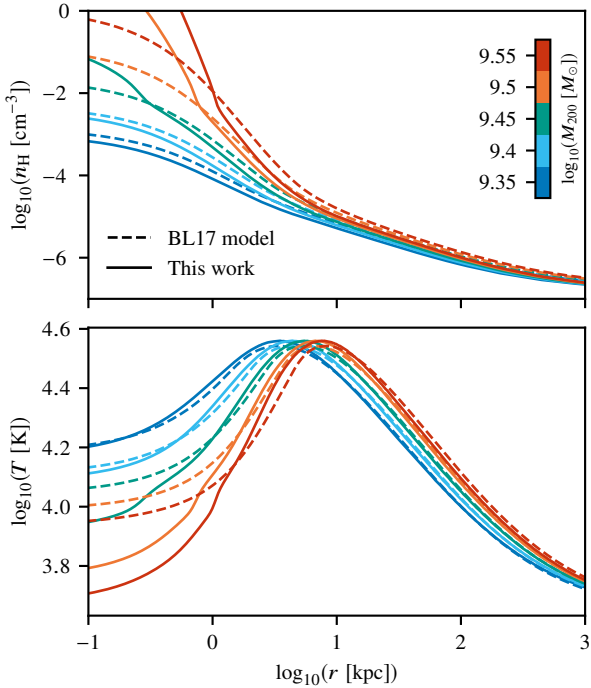}
\caption{Radial density (top) and temperature (bottom) profiles from \citetalias{benitez-llambayPropertiesDarkLCDM2017} model (dashed lines) and our code (solid lines). The models agree well, except for the inner regions of high-mass haloes, where the transition to neutral gas changes the mean mass per particle $\mu$, an effect which is not included in the \citetalias{benitez-llambayPropertiesDarkLCDM2017} model.}
\label{fig:relhic_comp}
\end{figure}

In addition to verifying the ionisation balance portion of our code, we compare the physical properties of our haloes to the analytic model of \citetalias{benitez-llambayPropertiesDarkLCDM2017}.
In Fig.~\ref{fig:relhic_comp}, we compare radial density and temperature profiles from this model, which is calibrated for the \citet{haardtRadiativeTransferClumpy2012} UVB model, and from our code run with the same spectrum.
We obtain good agreement for low-mass haloes and the outer regions of higher-mass haloes, where the gas remains fully ionised.
For haloes which are sufficiently massive to develop an ionisation front, our code predicts higher densities and lower temperatures within the neutral region.
This occurs because in the \citetalias{benitez-llambayPropertiesDarkLCDM2017} model, a constant value of the mean mass per particle $\mu=0.62$ is used, as appropriate for fully-ionised primordial gas.
In this work, we explicitly calculate the ionisation state of the gas as a function of radius, and so can determine $\mu$ as
\begin{equation}
\mu(r) = \frac{\sum_i X_i(r) A_i(r)}
		   {\sum_i X_i(r) + X_e(r)},
\label{eq:masspp}
\end{equation}
where $X_i$ is the number abundance of species $i$ relative to H, $A_i$ is the mass number of species $i$, and the $e$ subscript denotes electrons.
For fully neutral gas, Eq.~\ref{eq:masspp} yields $\mu=1.23$, resulting in higher pressures which require higher gas densities to maintain hydrostatic equilibrium.
Since the cooling rate is an increasing function of density, the equilibrium temperature for neutral gas is consequently lower.

\subsection{Star formation threshold}
\label{ssec:molfracs}

In the following section, we will show that more massive dark matter haloes will retain a more massive reservoir of gas; correspondingly, they will have larger neutral cores and higher peak values of $\Sigma_{\Ha}$.
However, these cores will be denser and colder, increasing the likelihood of star formation occurring.
UV radiation emitted locally from young stars within the halo will readily exceed the intensity of the UVB, while their supernovae would expel gas from the halo. Hence, we require that the total amount of star formation is small, which imposes an upper limit on the halo masses for which fluorescence is expected.

\interfootnotelinepenalty=10000

Star formation primarily occurs within dense molecular clouds, and so the presence of $\Hmol$ may be used as a proxy for star formation.
Ideally, we would self-consistently determine the molecular fraction within our code; however, the complexity of doing this puts it beyond the scope of this work.
Consequently, we instead consider the formation of $\Hmol$ as a post-processing step.
Analytic models, such as that described by \citet{krumholzStarFormationLaw2009}, permit this by prescribing the surface molecular fraction $f_{\Hmol} \equiv \Sigma_{\Hmol}/(\Sigma_{\HI} + \Sigma_{\Hmol})$ as a function of the column density of $\HI$ and an estimate of the $\Hmol$-dissociating Lyman-Werner flux.\footnote{More advanced models for $f_{\Hmol}$ exist, e.g. \citet{krumholzStarFormationLaw2013}. These model the formation of $\Hmol$ at low densities more accurately, but give the same results in the high-density regime that is relevant to our purposes.}
We note that this model is primarily intended for disk galaxies with active star formation and higher metallicity than the gas we consider (which we assume to have primordial composition, although for the purposes of computing $f_{\Hmol}$ we must assume a non-zero metallicity $Z=0.001Z_{\odot}$).
While the exact values of $f_{\Hmol}$ we obtain may not be fully accurate, we expect the column density threshold for $\Hmol$ formation to be reliable.
This threshold is predicted by the model to occur at $N_{\HI} \sim 10^{23}\,\text{cm}^{-2}$ which, when the absence of metals or dust is taken into consideration, is in good agreement with the observed paucity of damped $\Lya$ (DLA) systems above a similar column density threshold, believed to be due to the onset of $\Hmol$ formation \citep[e.g.][]{schayePhysicalUpperLimit2001,krumholzAbsenceHighMetallicityHigh2009}.
Additionally, the upper bound on halo mass of $\Mv\lesssim 10^{9.6}\,\Msun$ implied by this threshold is consistent with the fraction of `dark' haloes found in the \textsc{Apostle} hydrodynamical simulations, which exceeds 50\% for halo masses below this value \citep{benitez-llambayPropertiesDarkLCDM2017}.
Accordingly, excluding all haloes for which $N_{\HI} \geq 10^{23}\,\text{cm}^{-2}$ provides a threshold below which we may discount star formation.

An additional upper limit can arise from the constraint that the total gas fraction $M_{\text{gas}}/\Mv$ remains small, such that we are justified in neglecting the gas self-gravity.
At the halo masses required for this constraint to become relevant, we find that column densities already exceed the limit for molecular gas formation defined above.
However, the steep increase in gas density within the ionisation front raises the possibility of gas becoming locally self-gravitating within the neutral core.
Adopting the stricter condition $M_{\text{gas}}({\leq}\,r)/M_{\text{DM}}({\leq}\,r) \ll 1 \Forall r<\Rv$ to avoid this possibility would result in a small ($\lesssim 0.005\,\text{dex}$) reduction in our upper threshold on halo mass.
Nevertheless, in the absence of further work to model the chemical state of this gas, especially during early times when the greater mean density of the Universe would have yielded even steeper gas density profiles, our upper thresholds on halo mass should themselves be interpreted as upper limits.

\section{Properties of fluorescent haloes}
\label{sec:results}

\subsection{Fiducial model}

We first present results from our fiducial model, in which the UVB is given by the \citetalias{madauCosmicReionizationPlanck2015} spectrum and an NFW density profile is assumed for the dark matter halo, along with the \citet{ludlowMassConcentrationRedshift2016} mass-concentration relation and a redshift $z=0$.

\begin{figure}
\subimport{figs/}{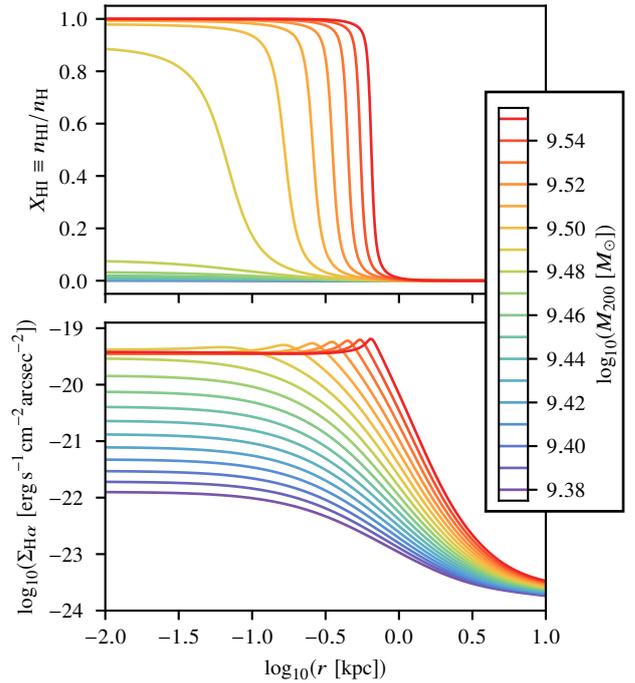}
\caption{Radial profiles for hydrogen neutral fraction (top) and $\Ha$ surface brightness (bottom) for haloes in our fiducial model, as a function of halo mass $\Mv$. Note the sudden transition in the ionisation state of the gas at $\Mv = 10^{9.5}\,\Msun$. This corresponds to the formation of the limb-brightened ring in emission, shown here as a peak in the $\Sigma_{\Ha}$ profile.}
\label{fig:profiles_fid}
\end{figure}

\begin{figure*}
\import{figs/}{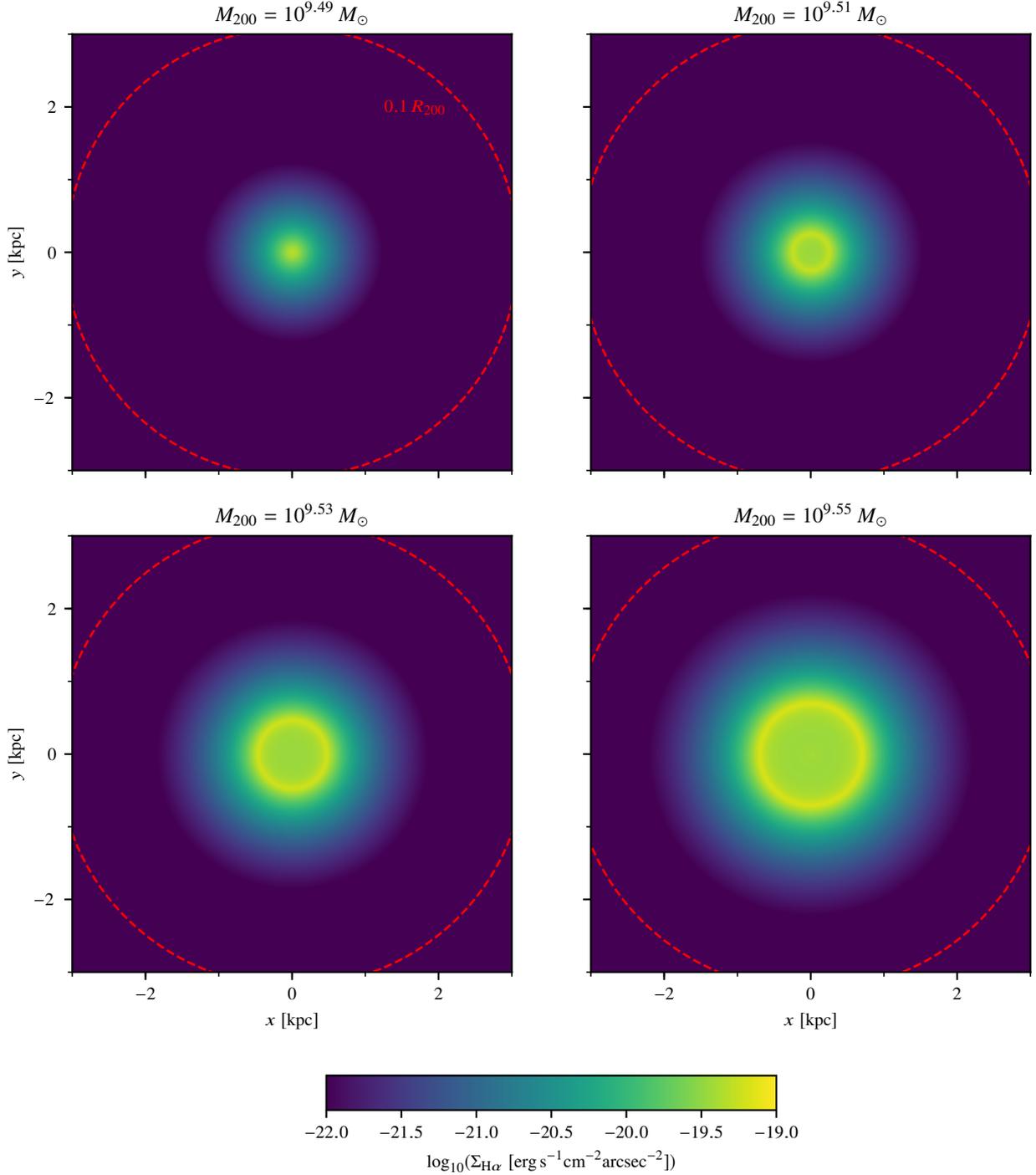}
\caption{Projected $\Ha$ surface brightness for haloes in our fiducial model. The range of halo masses shown is chosen such that in the top-left panel, a neutral core first begins to form. As the halo mass increases, the neutral core enlarges and the limb-brightened peak in $\Sigma_{\Ha}$ becomes more prominent until we reach the most massive halo allowed by our star formation criterion (bottom-right). Red dashed lines indicate 10\% of the virial radius.}
\label{fig:sb_contour_fid}
\end{figure*}

We compute models for haloes with masses $10^{9.38}\,\Msun$ and above, and show the resulting hydrogen neutral fractions $X_{\HI} \equiv n_{\HI} / n_{\text{H}}$ and $\Ha$ surface brightnesses $\Sigma_{\Ha}$ in Fig.~\ref{fig:profiles_fid}.
At the lower end of this range, hydrostatic equilibrium is achieved for a bound gas mass within $\Rv$ of $\text{M}_{\text{gas}} {\gtrsim} 10^7\,\Msun$.
With projected column densities $N_{\HI} \lesssim 10^{15}\,\text{cm}^{-2}$, this gas is optically thin to the UVB, and hence is highly ionised throughout.
The gas remains rarefied, with peak densities $n_{\text{H}} \sim 10^{-3}\,\text{cm}^{-3}$, and therefore $\Sigma_{\Ha}$ is low at all impact parameters.
As the halo mass increases, the central gas density also increases.
Consequently, at small impact parameters the projected column density and surface brightness gradually increase, while remaining largely unchanged along lines of sight avoiding the high-density central region.
For $\Mv = 10^{9.47}\,\Msun$, the central column density reaches $10^{18}\,\text{cm}^{-2}$, corresponding to an optical depth at the Lyman limit $\tau_{912} \sim 1$.
A sharp transition in the gas properties occurs as this threshold is exceeded: the gas becomes self-shielded from the UVB and develops a neutral core.
By $\Mv = 10^{9.49}\,\Msun$ the central neutral fraction approaches 1.

Fig.~\ref{fig:sb_contour_fid} shows the projected surface brightness distributions for haloes at this mass and above, clearly illustrating their characteristic ring-shaped surface brightness morphology.
This occurs because the $\Ha$ emissivity is sharply peaked at the ionisation front: at larger radii, the gas density is reduced, while inside the ionisation front few $\HII$ recombinations occur.
Hence, lines of sight tangent to the surface of the neutral core pass through a greater fraction of high-emissivity gas than those passing through the centre of the neutral region.
This `limb brightening' effect produces surface brightness profiles which peak at the projected radius of the ionisation front.
As the halo mass increases further, the core grows in size (see also Fig.~\ref{fig:profiles_fid}), and rapidly becomes cooler and denser, leading to the formation of molecular $\Hmol$, and our previously-discussed star formation criterion is exceeded for haloes above $10^{9.55}\,\Msun$.
For a halo of this mass, the surface brightness profile peaks at $r=0.70\,\text{kpc}$ with a value of $\Sigma_{\Ha{},\;\text{max}}=6.58\times10^{-20}\,\sbunit$.

Ring-shaped $\Ha$ surface brightness morphologies were also predicted by \citetalias{sternbergAtomicHydrogenGas2002}, with a comparable peak value of $\Sigma_{\Ha{},\;\text{max}} \simeq 4 \times 10^{-20}\,\sbunit$.
However, this is attained for a UVB spectrum with hydrogen photoionisation rate ${\simeq} 20\%$ lower than the \citetalias{madauCosmicReionizationPlanck2015} spectrum (see Section~\ref{ssec:alpha_uv} below).
Additionally, \citetalias{sternbergAtomicHydrogenGas2002} adopt an external bounding pressure, which acts in conjunction with the DM potential to confine the gas toward the centre of the halo, allowing a dense, self-shielding core to form at a much lower halo mass $\Mv \simeq 10^8\,\Msun$.

\begin{figure*}
\import{figs/}{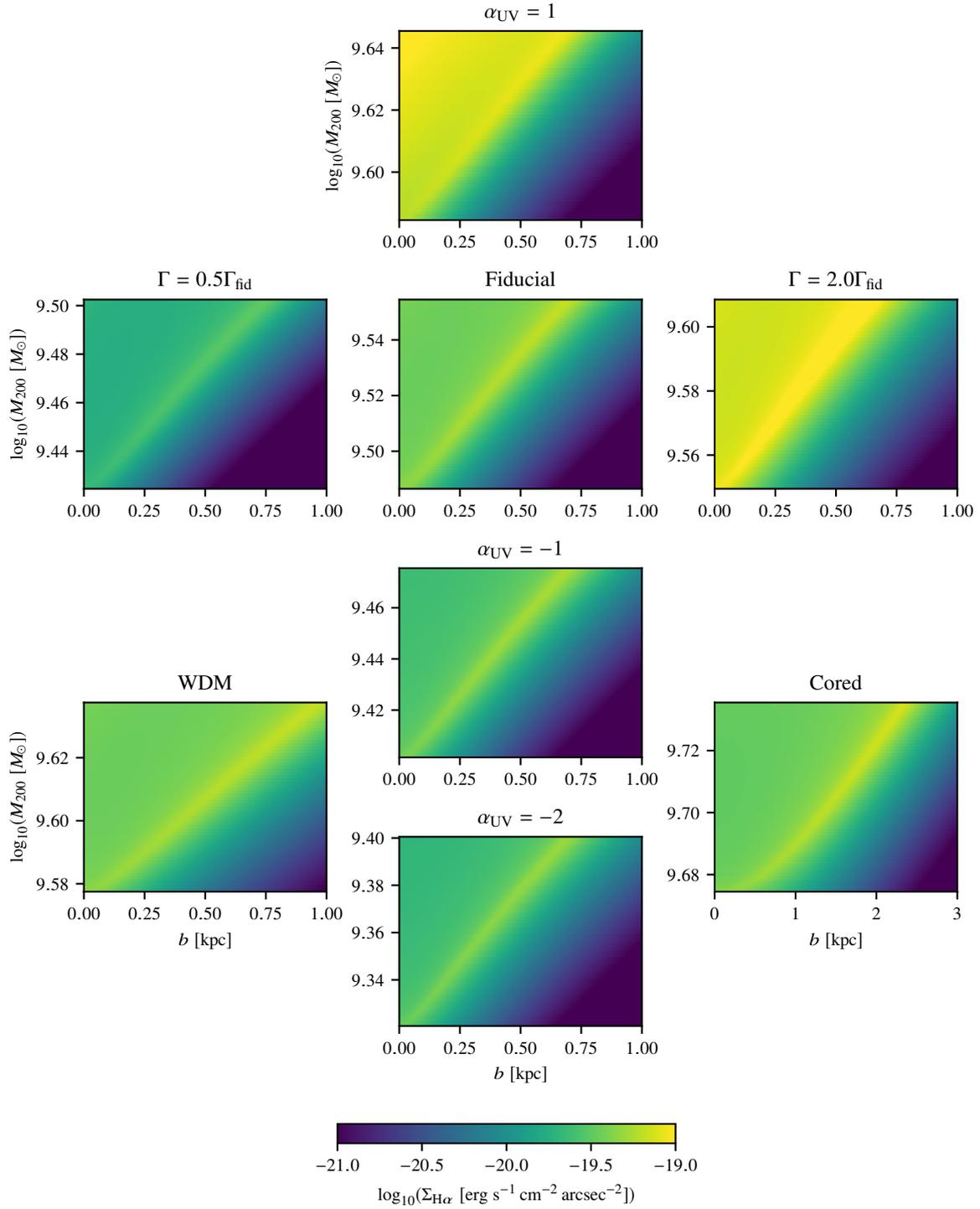}
\caption{$\Sigma_{\Ha}$ as a function of impact parameter $b$ and $\Mv$ for different model variations.
In each panel, the $y$-axis limits are set by the minimum mass required for the formation of a neutral core, and the maximum mass allowed by the star formation criterion. 
Results from our fiducial model are shown on the upper-central panel, with panels to the left and right illustrating the effect of varying the photoionisation rate, and panels above and below that of changing the spectral index $\alpha_{\text{UV}}$.
Panels in the bottom left and right corners show results for models with a different halo mass-concentration relation and halo density profile (note the wider $x$-axis limits in the latter).}
\label{fig:results_strips_all}
\end{figure*}

\subsection{Model variations}


\subsubsection{UVB spectrum}
\label{ssec:alpha_uv}

Discrete sources of ionising radiation, such as a nearby AGN or bright galaxy, can produce local inhomogeneities in the UVB.
Furthermore, uncertainties present in the modelling of the UVB result in variations in the intensity and spectral shape of the background computed by different models at a fixed redshift.
Accordingly, we explore the sensitivity of our results to such changes in the UVB.
The intensity of the UVB may be parameterised by the hydrogen photoionisation rate, which is given by
\begin{equation}
    \Gamma = 4 \pi \int_{\nu_{\text{th}}}^\infty \frac{J(\nu) \sigma(\nu)}{h \nu}\,\df{\nu},
    \label{eq:gamma0}
\end{equation}
where $J(\nu)$ is the mean intensity, $\sigma(\nu)$ is the cross-section for photoionisation of $\text{H}^0$, and $\nu_{\text{th}} = 13.6\,\text{eV} / h$.
In Fig.~\ref{fig:results_strips_all}, we show stacked radial surface brightness profiles as a function of halo mass, with colour encoding the value of $\Sigma_{\Ha}$.
The panel in the centre of the upper horizontal row corresponds to our fiducial model, in which the photoionisation rate is $\Gamma_{\text{fid}}=5.47 \times 10^{-14}\,\text{s}^{-1}$.
Panels to the left and right show results obtained with $\Gamma$ respectively decreased and increased by a factor of two from its fiducial value.
We find that increasing the photoionisation rate shifts the halo mass scale at which fluorescent rings are predicted upward, while the surface brightness of these rings also increases.
Conversely, a reduced photoionisation rate allows dimmer rings to form at lower halo masses.

We interpret these trends as being due to the ability of a more intense UVB to penetrate a greater total column density of gas (i.e. $N_{\text{H}}$) before becoming attenuated.
In the situation we consider, the gas volume density is a function of radius, which itself becomes steeper with increasing halo mass.
This results in the ionisation front forming at a larger characteristic gas number density, which is only reached for more massive haloes.
In turn, the greater density of the gas at the ionisation front boosts its emissivity, and hence the peak surface brightness.
We further explore the reasons for surface brightness variations between models in Section~\ref{sec:discuss}.

\begin{figure}
\subimport{figs/}{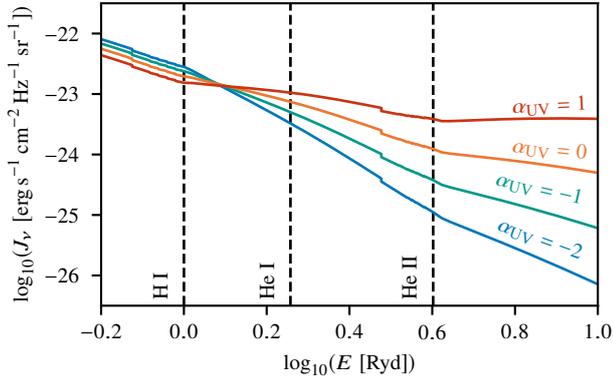}
\caption{UVB spectra in the range $1\;\text{Ryd} < E < 10\;\text{Ryd}$ for different values of the slope parameter $\alpha_{\text{UV}}$, where $\alpha_{\text{UV}}=0$ corresponds to the fiducial \citetalias{madauCosmicReionizationPlanck2015} spectrum. Vertical dashed lines indicate the $\HI$, $\HeI$ and $\HeII$ ionisation edges.}
\label{fig:alpha_uv_spectra}
\end{figure}

To consider variations in the shape of the UVB spectrum, we adopt the parameterisation from \citet{crightonMetalenrichedSubkiloparsecGas2015}, which introduces the UVB shape parameter $\alpha_{\rm{UV}}$.
This parameter is used to modify the ionising spectrum as follows:
\begin{equation}
J_{\nu}(E)=\begin{cases}
N_{\Gamma} \times H(E) & E \leq E_0\\
N_{\Gamma} \times H(E) \times (E/E_0)^{\alpha_{\rm{UV}}} & E_0 < E \leq E_1\\
N_{\Gamma} \times H(E) \times (E_1/E_0)^{\alpha_{\rm{UV}}} & E > E_1,
\end{cases}
\end{equation}
where $H(E)\equiv J_{\nu,\text{MH15}}(E)$, $E_0=1\,\text{Ryd}$ and $E_1=10\,\text{Ryd}$.
We additionally scale these spectra by a factor $N_{\Gamma}\equiv\Gamma_{\text{fid}}/\Gamma$, such that the photoionisation rate is unchanged from its fiducial value, in order to examine the effect of modifying the spectral shape in isolation.
The resulting spectra for $\alpha_{\text{UV}} = 0$ (corresponding to our fiducial model), ${+}1$, $-1$ and $-2$ are plotted in Fig.~\ref{fig:alpha_uv_spectra}.

In Fig.~\ref{fig:results_strips_all}, the central column shows $\Ha$ surface brightness profiles as a function of $\Mv$ for haloes illuminated by the same four spectra.
We find that the halo mass required to form a fluorescent ring is reduced for more negative values of $\alpha_{\text{UV}}$, corresponding to softer spectra with a greater proportion of ionising photons having energies close to the Lyman limit (since the spectra have been renormalised to maintain a constant value of $\Gamma$).
These photons are more likely to be absorbed, due to the ${\sim}\nu^{-3}$ dependence of the photoionisation cross-section, and so will on average be absorbed within a lower total gas column density.
As described previously, the coupling between column and volume densities means that this corresponds to the ionisation front forming in lower-density gas, which maintains hydrostatic equilibrium with a shallower dark matter potential.
In addition to the reduction in emissivity due to the lower gas density, an additional effect is caused by the reduced flux of photons with energies close to $24.6\;\text{eV}$, at which helium is singly ionised.
This produces a larger $\HeI$ fraction at the H ionisation front, further reducing the emissivity by lowering the electron density.
Opposite arguments apply to the harder $\alpha_{\text{UV}}=1$ spectrum, which produces larger peak surface brightnesses but inhibits neutral cores from forming until higher halo masses are reached.

\subsubsection{Halo density profile}
\label{ssec:dmprof}

In our fiducial model, we have used an NFW density profile to model dark matter haloes, with the concentration parameter $c_{200}$ determined using the \citet{ludlowMassConcentrationRedshift2016} relation, which is obtained from haloes in the Copernicus Complexio (\textsc{CoCo}) N-body simulations \citep{boseCopernicusComplexioStatistical2016,hellwingCopernicusComplexioHighresolution2016}.
The \textsc{CoCo} suite also includes runs in which a warm dark matter (WDM) cosmology is used. In contrast to CDM cosmologies, where $c_{200}$ is a monotonically increasing function of $\Mv$, the WDM mass-concentration relation turns over at a specific mass scale~\citep{boseSubstructureGalaxyFormation2017}.
$\Ha$ surface brightness profiles obtained by using this relation are shown in the bottom-left corner of Fig.~\ref{fig:results_strips_all}.
At the halo masses at which we predict fluorescent ring emission, the WDM mass-concentration relation results in haloes which are approximately 50\% less concentrated at fixed $\Mv$ than the equivalent halo following the L16 relation, and so have systematically lower central dark matter densities.
Consequently, hydrostatic equilibrium is maintained with a shallower gas density profile, which produces lower central gas densities.
Hence, the column density required to form a neutral core is first reached at higher halo mass, and once neutral cores form they extend to larger projected radii.

We also consider the possibility of varying the density profile assumed for the dark matter, by computing surface brightness profiles for haloes with `cored' density profiles, where the central dark matter density approaches a finite value rather than diverging to infinity, as occurs for `cuspy' profiles such as NFW.
Observations of nearby dwarf galaxies, which imply central density gradients shallow than an NFW profile, have been interpreted either as a manifestation of the self-interacting nature of dark matter \citep[e.g.][]{spergelObservationalEvidenceSelfInteracting2000}, or as a result of baryonic interactions \citep[][and references therein]{navarroCoresDwarfGalaxy1996,pontzenHowSupernovaFeedback2012}.
The latter scenario generally requires supernova feedback from stars forming within the halo \citep[e.g.][]{benitez-llambayBaryoninducedDarkMatter2018}, and so evidence of a core in a halo hosting a fluorescent ring would be suggestive of self-interacting dark matter (SIDM).
We choose to model Burkert density profiles, for which \citep{burkertStructureDarkMatter1995}:
\begin{equation}
\rho(r)=\frac{\rho_0 {r_0}^3}{(r+r_0)(r^2+{r_0}^2)},
\label{eq:dp_burkert}
\end{equation}
where $r_0$ is the core radius, which we choose to be a function of $\Mv$ such that $\Rv / r_0 = c_{200}$, in analogy to the NFW scale radius $r_s$.
This choice results in core radii $r_0 \sim 2 - 3\,\text{kpc}$, somewhat larger than the typical size $r_0 \sim 1\,\text{kpc}$ predicted for SIDM haloes of the mass we consider, but within the range allowed by constraints on the self-interaction (see e.g. \citealt{tulinDarkMatterSelfinteractions2018}).
We caution that these assumptions will influence the resulting fluorescent emission; accordingly, the results we present here, shown in the bottom-right panel of Fig.~\ref{fig:results_strips_all}, should be treated as illustrative only, rather than a specific prediction of a particular SIDM model.
As with the \citet{boseSubstructureGalaxyFormation2017} mass-concentration relation, the reduction in central dark matter density caused by the dark matter core results in larger, more extended neutral gas cores developing at higher halo mass.
However, the more extreme modification to the dark matter density profile results in a much more substantial change in the size of the neutral gas core.
In both these models and those using the WDM mass-concentration relation, we find the magnitude of $\Sigma_{\Ha{},\text{max}}$ to be slightly increased with respect to our fiducial model.
This occurs because of the larger physical size of the emitting region results in a longer path length over which $\varepsilon_{\Ha}$ is integrated to obtain the surface brightness, but is a relatively minor effect when compared to the impact of varying the properties of the UVB.

\section{Detectability of fluorescent haloes}
\label{sec:abund}

\begin{figure}
\subimport{figs/}{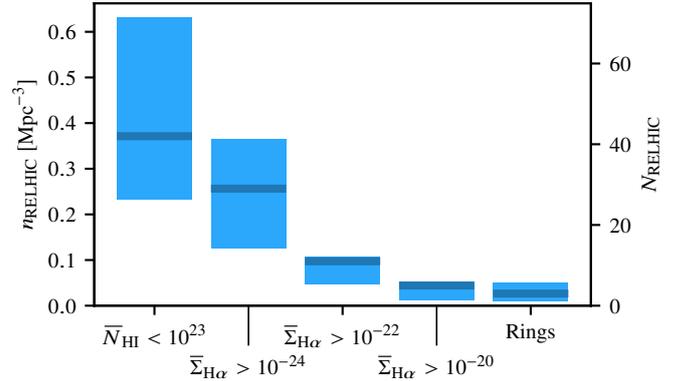}
\caption{Number density distribution of $\Ha$-fluorescent haloes in \textsc{apostle}. The median values are shown by thick solid lines; shaded regions indicate $10^{\text{th}}-90^{\text{th}}$ percentiles. From left to right, we show the number density of all RELHICs with $\HI$ column densities below our star formation threshold, those with surface brightnesses exceeding three different limiting values, and those with discernible ring-shaped surface brightness profiles. Here, $\protect\widebar{N}_{\HI} \equiv N_{\HI} / [\text{cm}^2]$ and $\protect\widebar{\Sigma}_{\Ha} \equiv \Sigma_{\Ha} / [\sbunit]$. The right-hand axis indicates the corresponding number counts per \textsc{apostle} volume.}
\label{fig:relhic_distrib}
\end{figure}

We now make predictions for the abundance with which dark haloes exhibiting $\Ha$ fluorescence appear on the sky, and hence consider the prospects for their detection, using results from the \textsc{apostle} suite of high-resolution hydrodynamical simulations~\citep{sawalaAPOSTLESimulationsSolutions2016}.
These simulations follow twelve galaxy groups selected to resemble the Local Group (LG) on the basis of their dynamical properties; we consider the five volumes simulated at the highest (`L1') resolution corresponding to a gas (DM) particle mass $1.0\ (5.0) \times 10^4\,\Msun$.
From the $z=0$ outputs of these simulations, we select all haloes with $M_{200} > 10^8\,\Msun$ located within 3.5\,Mpc of the simulated LG's barycentre, before discarding all `luminous' haloes (i.e. those containing one or more star particles).
The remaining `dark' haloes are then divided into the two populations (COSWEBs and RELHICs) introduced in Section~\ref{sec:intro}, where haloes with negligible gas content at $z=0$ are placed into the former category and discounted.\footnote{The group finder \textsc{subfind}~\citep{springelPopulatingClusterGalaxies2001} is used to quantify the bound gas mass.}
From this classification, we obtain a total of 263 RELHICs across the five volumes, with median halo mass $M_{200}=10^{9.12}\,\Msun$.
We match each of these haloes with a corresponding calculation from Section~\ref{sec:results}, deselecting 23 objects due to their exceeding our star formation criterion (Section~\ref{ssec:molfracs}); for the remaining 240 we obtain a prediction of their peak $\Ha$ surface brightness under our fiducial model.
In Fig.~\ref{fig:relhic_distrib}, we show the resulting number density of all fluorescent objects, and of subsets with surface brightnesses exceeding chosen threshold values.

Fig.~\ref{fig:relhic_distrib} indicates that high surface brightness fluorescence is rare, as are ring-shaped structures, with a median value of $3^{+2.6}_{-2.0}$ objects per volume, where the sub- and superscripts indicate the 10th and 90th percentile limits.
This is not unexpected, due to the very narrow range in halo mass for which we expect bright ring fluorescence.\footnote{We caution that `bright' is used here in a relative sense only; in absolute terms a surface brightness ${\sim}10^{-20}\,\sbunit$ is still observationally challenging. Nevertheless, a combination of deep observations, high spectral resolution, and pixel-based stacking techniques has been successfully used to image surface brightnesses approaching this threshold~\citep[e.g.][]{fumagalliMeasurementUVBackground2017,wisotzkiNearlyAllSky2018}.}
Consequently, detecting fluorescent rings in the surroundings of the LG is impractical, since a prohibitively deep, all-sky $\Ha$ survey would be required.
However, searching for fluorescent haloes around nearby galaxies, albeit still challenging, becomes more feasible, since surface brightness remains independent of distance for cosmologically-nearby sources, but the angular extent of a volume equivalent to that simulated in \textsc{apostle} decreases substantially.
In this situation, ultra-deep observations along single pointings have the potential to uncover one or more florescent rings.

Additionally, the requirement for an expensive optical search may be circumvented by exploiting the large $\HI$ column densities that will be associated with strongly-fluorescent systems.
These imply that haloes hosting fluorescent rings should also be bright 21\,cm $\HI$ emitters, and so existing $\HI$ sources may be used to provide a set of targets for subsequent $\Ha$ observations, once deep (but affordable) broad-band imaging has been used to rule out associations with galaxies.
Of particular interest is the catalogue of ultra-compact high velocity clouds (UCHVCs) detected in surveys like ALFALFA \citep{adamsCatalogUltracompactHigh2013}, which \citetalias{benitez-llambayPropertiesDarkLCDM2017} demonstrate to have $\HI$ emission properties which overlap at least partially with corresponding predictions for \textsc{apostle} RELHICs.
This is in contrast to the (non `ultra-compact') HVCs, given the `extragalactic' environment for RELHICs that \textsc{apostle} favours. \citetalias{sternbergAtomicHydrogenGas2002} found that the latter scenario would require HVCs to be underconcentrated and unrealistically spatially extended in order to reproduce their observed angular sizes.

\begin{figure}
\vspace{-\baselineskip}
\subimport{figs/}{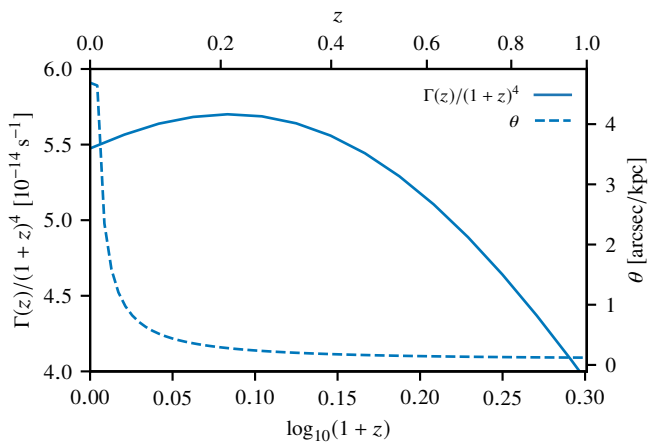}
\caption{Redshift evolution of the apparent surface brightness (solid curve; left scale) and the angular separation corresponding to a comoving kpc (dashed curve; right scale).}
\label{fig:z_sweet}
\end{figure}

Throughout the preceding discussion, we have considered fluorescence from haloes at redshift $z=0$.
At earlier times, the UVB has a greater intensity, resulting in an increased photoionisation rate $\Gamma$.
As detailed in Section~\ref{ssec:alpha_uv}, this will correspond to greater intrinsic surface brightnesses.
However, the observed surface brightness is dimmed by a factor $1/(1+z)^4$ relative to its intrinsic value.
In Fig.~\ref{fig:z_sweet}, we illustrate that the quantity $\Gamma(z) / (1+z)^4$, which is directly proportional to the observed surface brightness, is maximised at $z \sim 0.2$.
Also shown in Fig.~\ref{fig:z_sweet} is the variation with redshift of the angular separation corresponding to a proper kpc, which decreases rapidly with redshift.
In addition, halo concentration increases with redshift, which we expect to result in more compact fluorescent rings.
Hence, resolving fluorescent objects at $z \gg 0$ is likely to be challenging.
As we noted in Section~\ref{ssec:alpha_uv}, a boost in the UVB photoionisation rate can be provided even at $z \simeq 0$ by proximity to a highly-luminous source, such as an AGN \citep{cantalupoFluorescentLyUpalpha2005}.
To obtain detailed predictions of the expected surface brightness in this case, further modelling is required to determine the precise effects of the resulting change in spectral shape on the properties of fluorescent rings.
Additionally, hydrodynamic modelling is needed to establish the conditions (e.g. separation from the source) needed to avoid stripping of gas from a candidate halo.

\section{Physics governing the properties of fluorescent haloes}
\label{sec:discuss}

\begin{figure}
\subimport{figs/}{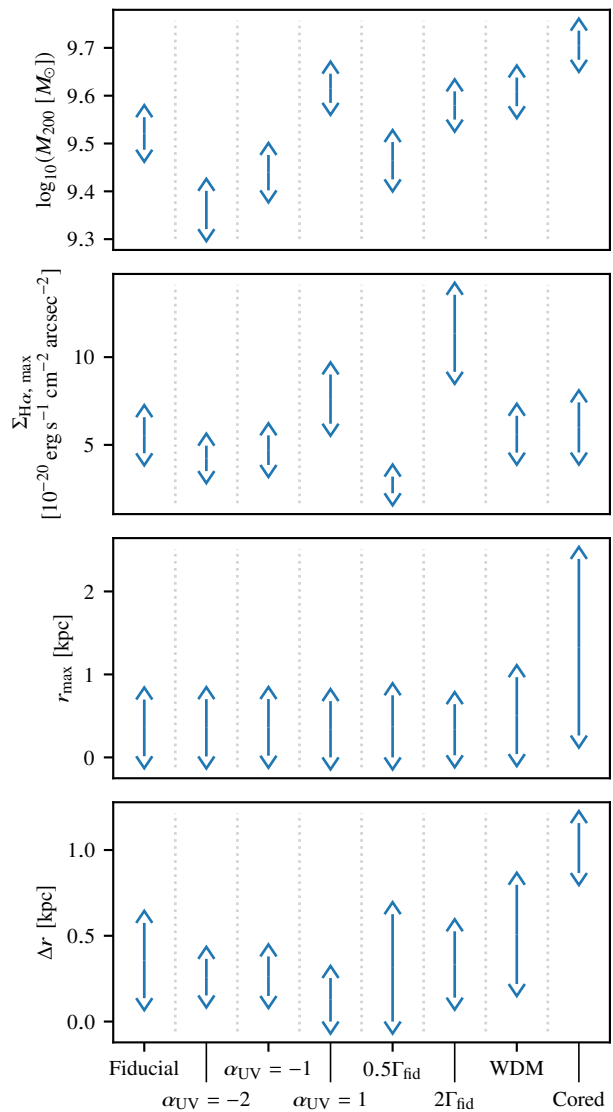}
\caption{Summary of properties of ring fluorescence. The top panel shows the halo mass range for which ring fluorescence is predicted, for all models considered in Section~\ref{sec:results}.
Our fiducial model is on the far left, followed by the following variations: three different values of the UVB spectral slope $\alpha_{\text{UV}}$, UVB photoionisation rates $\Gamma$ a factor of 2 below and above the fiducial value, the WDM \citep{boseSubstructureGalaxyFormation2017} mass-concentration relation, and the cored dark matter density profile given by \citet{burkertStructureDarkMatter1995}.
Subsequent panels show, for the same set of haloes in each model, values of the peak surface brightness, the radius at which this peak is reached, and the full width at half maximum of the radial surface brightness.}
\label{fig:results_whiskers}
\end{figure}

In Section~\ref{sec:results}, we presented surface brightness profiles resulting from varying input parameters to our photoionisation code.
In the event of a successful detection of an object with ring-shaped fluorescent $\Ha$ emission, we would like to determine the physical properties of the halo hosting the ring along with the incident radiation field.
Hence, in this section we consider the causes of the trends displayed in our results in more detail, and discuss how they may be used to infer the properties of a fluorescent ring and its environment from its appearance.

We summarise our results in Fig.~\ref{fig:results_whiskers}, which shows the range of masses for which ring fluorescence is predicted in the different model variations we consider, along with the corresponding ranges in the peak value surface brightness, the projected radius at which this peak occurs, and the full width at half maximum of the surface brightness profile.
Depending on its properties, purely detecting a fluorescent ring may provide constraints on its environment.
For example, detection of a ring with $\Sigma_{\Ha,\,\text{max}} \gtrsim 10^{-19} \, \sbunit$ would strongly suggest a UVB intensity enhanced with respect to the \citetalias{madauCosmicReionizationPlanck2015} spectrum.
From the bottom two panels of Fig.~\ref{fig:results_whiskers}, we also see that with `non-standard' dark matter (i.e. WDM or cored density profiles), fluorescent rings are expected to reach their peak surface brightness at larger projected radii than in the fiducial cuspy CDM case. However, the surface brightness profiles for the cored haloes we model are also broader, meaning that combined measurements of $r_{\text{max}}$ and $\Delta{r}$ could potentially be used to distinguish WDM from CDM, and cored from cuspy density profiles.
We again note that these properties depend on the exact parameterisation of the density profile.
For example, choosing a smaller core radius, corresponding to a more concentrated halo, would translate to reductions in $r_{\text{max}}$ and $\Delta r$.

\begin{figure}
\subimport{figs/}{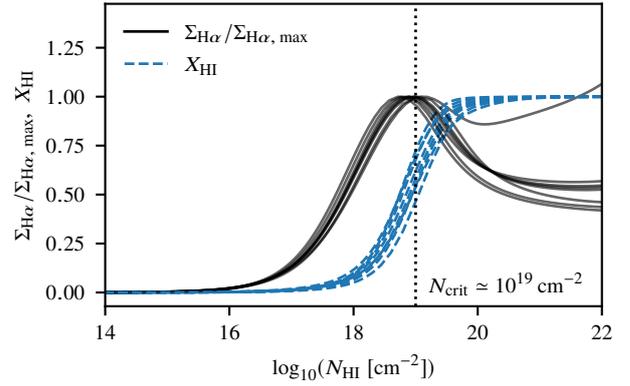}
\caption{Normalised $\Ha$ surface brightness profiles for the most massive halo satisfying our star formation criterion in each model we consider (black), shown with corresponding H neutral fractions $X_{\HI}$ (dashed blue).}
\label{fig:sb_stack_all}
\end{figure}

We now turn to considering why different input parameters affect the properties of fluorescent rings in the ways shown in Fig.~\ref{fig:results_whiskers}.
An important trait universal to all our models is that, despite significant variation in the sizes and peak surface brightnesses of the fluorescent rings they predict, the column density of the gas at $r_{\text{max}}$ has a characteristic value $N_{\HI} = N_{\text{crit}} \simeq 10^{19}\,\text{cm}^{-2}$.
We illustrate this in Fig~\ref{fig:sb_stack_all}, where we also show that this corresponds to the column density at which the neutral fraction is $X_{\HI} = 0.5$.
While the association of the ionisation front with a column density of this magnitude is a standard result of radiative transfer calculations, it is nevertheless significant that we obtain self-similar $X_{\HI}$ and $\Sigma_{\Ha}$ profiles despite the substantial variation in the shapes of physical density profiles between our different models.

One notable exception to this self-similarity is the $\alpha_{\text{UV}}=1$ model, in which the surface brightness profile dips after the peak at $N_{\text{crit}}$, but then rises again toward higher $N_{\HI}$.
This occurs due to the increased flux of high-energy photons in the harder UVB spectrum assumed by this model, which cause ionisations at a larger typical value of column density (corresponding to smaller radii) due to their reduced photoionisation cross-section.
Additionally, the energetic photoelectrons they liberate can go on to ionise further atoms via secondary collisional processes.
This additional source of ionisation produces a greater residual ionised fraction within the ionisation front. In combination with the increase in $n_{\text{H}}$ toward smaller radii, this produces the observed doubly-peaked surface brightness profile.

With this `fixed point' identified, the changes in ring brightness and radius between models may be described in terms of changing $N_{\HI}-n_{\text{H}}$ and $N_{\HI}-r$ relations respectively.
The former follows since the peak emissivity is given by $\varepsilon_{\Ha,\,\text{max}}=\left.X_{\HII}n_{\text{H}}n_{\text{e}}\alpha_{\Ha}(T)\right|_{r_{\text{max}}}$.
Since $r_{\text{max}}$ corresponds to the ionisation front where $X_{\HII} = 0.5$, and so $n_{\text{e}} \approx n_{\text{H}}$ while also $T \simeq 10^4\,\text{K}$, the value of $\varepsilon_{\Ha,\,\text{max}}$ is set by $n_{\text{H}}(r_{\text{max}})$. 
Hence, a change in the local density of gas where $N_{\HI}$ reaches $N_{\text{crit}}$ will produce a change in the peak surface brightness.
By the same token, a change in $r_{\text{max}}$ implies a change in the position of the ionisation front, and therefore a change in the radius at which $N_{\HI}=N_{\text{crit}}$.
With these `scaling relations' identified, it is then possible to take the properties of an observed fluorescent ring and work backwards to determine the physical properties of the host halo.
By comparing these inferred values with those predicted by our simulations, quantitative constraints may then be placed on which models are applicable to observed objects, as we discuss next.

\begin{figure}
\subimport{figs/}{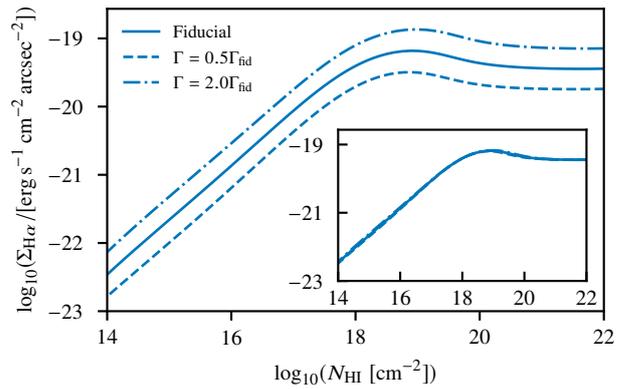}
\caption{Surface brightness as a function of $\HI$ column density for different photoionisation rates. In the inset panel, we show that the three $N_{\HI}$--$\Sigma_{\Ha}$ relations are self-similar when the surface brightness is scaled by the ratio $f_\Gamma = \Gamma / \Gamma_{\text{fid}}$.}
\label{fig:gamma_selfsim}
\end{figure}

We first consider trends in the maximum surface brightness $\Sigma_{\Ha,\,\text{max}}$; as described previously, this is primarily set by the gas number density at the ionisation front.
The physical size of the neutral region provides an additional influence, by altering the path length of sightlines passing through gas with emissivity close to maximal.
However, the most extreme change in ring geometry with respect to our fiducial model occurs when a cored density profile is assumed, but the resulting values of $\Sigma_{\Ha,\,\text{max}}$ differ by a much smaller margin.
Hence, we infer that the ring brightness is set primarily by properties of the UVB.
In particular, the models in which we consider a doubling or halving of $\Gamma$ predict changes in surface brightness by the same amount.
This is shown explicitly by plotting $\Sigma_{\Ha,\,\text{max}}$ as a function of $N_{\HI}$ (Fig.~\ref{fig:gamma_selfsim}), where we demonstrate in the inset panel that $\Sigma_{\Ha}/f_{\Gamma}$ curves, where $f_\Gamma$ is the multiplicative factor applied to the photoionisation rate, are self-similar.
Hence, assuming the other properties of the UVB (i.e. $\alpha_{\text{UV}}$) are similar to the MH15 spectrum, measuring the peak surface brightness of an observed ring and inferring a value for $f_{\Gamma}$ would allow a direct determination of the UVB photoionisation rate (see also \citealt{fumagalliMeasurementUVBackground2017}).

\begin{figure}
\subimport{figs/}{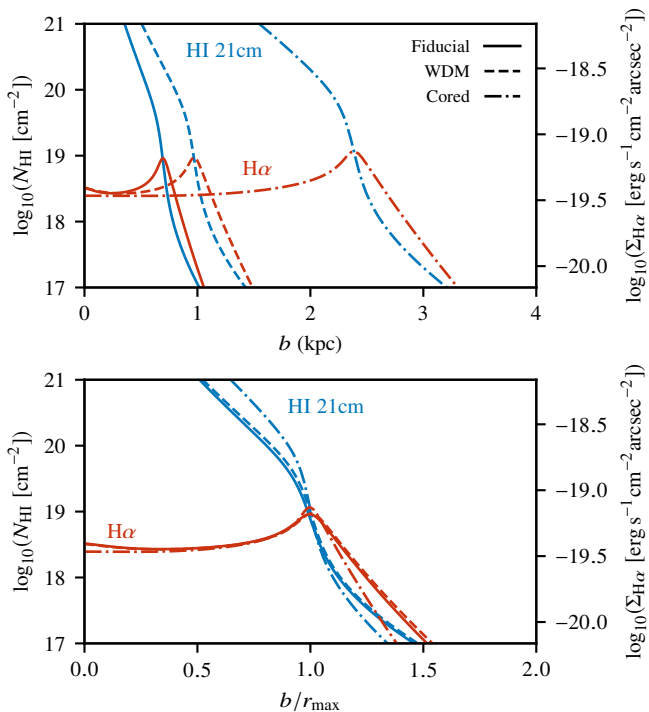}
\caption{$\HI$ column density profiles (blue) for models with different dark matter density profiles, with corresponding $\Ha$ surface brightness profiles (red). Profiles are shown as a function of impact parameter $b$, both in physical units (top panel) and normalised by $r_{\text{max}}$ (bottom panel). For each model, profiles are shown for the most massive halo which satisfies our star formation criterion.}
\label{fig:dmprof_selfsim}
\end{figure}

\begin{figure}
\subimport{figs/}{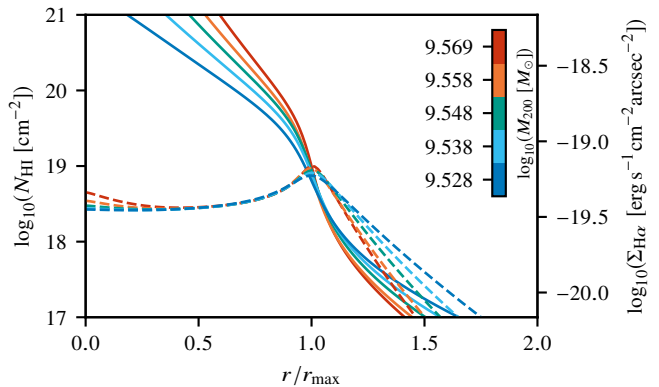}
\caption{$\HI$ column density profiles (solid lines), with corresponding $\Ha$ surface brightness profiles (red). Profiles are shown as a function of $b/r_{\text{max}}$ for haloes of different mass in our fiducial model.}
\label{fig:dmprof_onemodel}
\end{figure}

In contrast, the projected radius $r_{\text{max}}$ at which fluorescent rings are maximally bright is essentially independent of the properties of the UVB, changing only when the mass-concentration relation or density profile of the underlying dark matter halo is altered.
As the column density at $r_{\text{max}}$ remains equal to $N_{\text{crit}}$, it is the $N_{\HI}$--$r$ relation that is changing in this case, as we illustrate in the upper panel of Fig.~\ref{fig:dmprof_selfsim}.
Identifying this trend reinforces our previous observation that measuring $r_{\text{max}}$ in combination with $\Delta r$ permits cuspy CDM, cored DM (e.g. SIDM) and WDM to be differentiated.
However, doing so requires knowledge of the distance to the ring in order to determine physical values of $r_{\text{max}}$ and $\Delta r$. Without this, the $\Ha$ and $\HI$ profiles are only known as a function of $b/r_{\text{max}}$, a situation which results in them becoming highly self-similar.
We illustrate this in the lower panel of Fig.~\ref{fig:dmprof_selfsim}.
However, if we instead assume a fixed family of models and vary the halo mass, we find a systematic trend in the shape of the $\HI$ profile, which becomes steeper for $N_{\HI}>N_{\text{crit}}$ with increasing halo mass (Fig.~\ref{fig:dmprof_onemodel}).
Changes also occur in the $\Ha$ profiles, which become more sharply peaked at $b=r_{\text{max}}$ for higher halo masses.
Using complementary observations of $\Ha$ and $\HI$ in this way may permit better constraints to be made, although the relatively subtle change in e.g. $r_{\text{max}}$ produced by the WDM models, which correspond to a significant $~50\%$ reduction in halo concentration $c_{200}$, imply that obtaining the necessary accuracy from observations would be challenging.

\section{Summary and conclusions}

We have reported the results from radiative transfer calculations which allow the simulation of gas bound to a gravitational potential such as that due to a dark matter halo and illuminated by an isotropic ionising background.
By comparison with the well-known \textsc{cloudy} software, as well as previous analytic modelling, we have established that for the intended usage, despite making a number of simplifications, the code we use is robust and its results accurate.
We have used this code to model the gas clouds bound to low-mass dark matter haloes, and obtained surface brightness profiles for the $\Ha$ line, which is produced via fluorescent emission due to the ionising background.
Our key findings are:

\begin{enumerate}
    \renewcommand\theenumi{(\roman{enumi})}
    \item We confirm the results of previous studies, finding that haloes which develop a self-shielded neutral core possess surface brightness profiles with a characteristic ring-shaped morphology.
    \item This occurs for only a narrow range of halo masses, bounded from below by the requirement of a sufficiently deep potential, and from above by the requirement that star formation remains suppressed.
    \item Using numerical simulations, we predict a small population of $3^{+2.6}_{-2.0}$ (10-90\% c. i.) fluorescent rings in a spherical 3\,Mpc volume centred on the Local Group. Detecting these objects via a blind $\Ha$ survey would be impractical, but pre-selection techniques, e.g. via 21\,cm $\HI$ imaging, can provide catalogues of promising target sources, particularly beyond the LG where the sky coverage required to map a region of equivalent volume is reduced.
    \item We quantify how variations of the intensity (i.e. photoionisation rate) and spectral shape of the ionising background map directly to changes in the intensity of the fluorescent emission. Modifications to the dark matter potential instead alter the apparent size and shape of the fluorescent ring.
    \item We show that reverse-engineering these relations would enable a direct determination of the UVB photoionisation rate from the intrinsic brightness of an observed ring; equally, constraints on the nature and distribution of dark matter may be inferred from the ring's geometry.
\end{enumerate}

The existence of a large number of low-mass dark matter halos which do not host a luminous galaxy is an unavoidable prediction of the CDM paradigm.
While challenging, observing the distinctive ring-shaped fluorescent $\Ha$ emission from the gas bound to these haloes would provide a probe of this cosmological model on an as-yet untested scale.
In the longer term, a detailed analysis of this emission would unlock further insights, both into the nature of dark matter and the properties of the extragalactic ionising background.

\section*{Acknowledgements}

CS acknowledges support by a Science and Technology Facilities Council studentship [grant number ST/R504725/1].
MF \& TT acknowledge support by the Science and Technology Facilities Council [grant number ST/P000541/1]. This project has received funding from the European Research Council (ERC) under the European Union's Horizon 2020 research and innovation programme (grant agreement No. 757535).
During this work, RJC was supported by a Royal Society University Research Fellowship.
We thank the members of the \textsc{apostle} project for providing access to data.
This work has benefited from the public Python packages \textsc{numpy}, \textsc{scipy}, \textsc{astropy} and \textsc{matplotlib}.
This work used the DiRAC Data Centric system at Durham University, operated by the Institute for Computational Cosmology on behalf of the STFC DiRAC HPC Facility (www.dirac.ac.uk). This equipment was funded by BIS National E-infrastructure capital grant ST/K00042X/1, STFC capital grant ST/H008519/1, and STFC DiRAC Operations grant ST/K003267/1 and Durham University. DiRAC is part of the National E-Infrastructure.




\bibliographystyle{mnras}
\bibliography{sph_cloudy_paper} 




\appendix

\section{Analytic estimate of \texorpdfstring{$\Ha$}{H-alpha} surface brightness in optically-thick limit}
\label{app:ppcase}
\newcommand{\sigavg}{\ensuremath\langle\sigma\rangle}
In this Appendix we present a derivation of an analytic result for the $\Ha$ surface brightness in the optically-thick limit (see Fig.~\ref{fig:sb_noHe_GW}), following the method discussed by \citet{gouldImagingForestLyman1996}.
For a constant-density, plane-parallel slab of pure hydrogen composition, photoionisation equilibrium gives
\begin{equation}
    \alpha(1-X_{\HI})^2 n_{\rm{H}}^2 = \Gamma X_{\HI} n_{\rm{H}}
    \label{eq:gw1}
\end{equation}
where $X_{\HI} = n_{\HI} / n_{\rm{H}}$, and $\alpha$ is the recombination coefficient for $H^{+}$.
At a depth $\ell$ inside the cloud, the photoionisation rate is given by its value at the illuminated face of the cloud $\Gamma_0$, attenuated by the total optical depth $\tau(\ell)$, i.e.
\begin{equation}
    \Gamma(\ell) = \Gamma_0 \exp\left[-\tau(\ell)\right]
    \label{eq:gw2}
\end{equation}
The $\Lya$ surface brightness in units of $\text{photons}\;\text{cm}^{-2}\;\text{s}^{-1}$ is given by
\begin{equation}
    \Sigma_{\Lya} = f_{\Lya} \int_0^\ell \alpha(1-X_{\HI})^2 n_{\rm{H}}^2\,\df{\ell'}
    \label{eq:gw3}
\end{equation}
where $f_{\Lya} \sim 0.6$ is the fraction of recombinations that lead to emission of a $\Lya$ photon \citep{osterbrockAstrophysicsGaseousNebulae2006}, and $\alpha$ is the recombination coefficient.
Substituting Eqs.~\ref{eq:gw1} and \ref{eq:gw2} into the above then gives
\begin{equation}
    \Sigma_{\Lya} = f_{\Lya} \int_0^\ell X_{\HI} n_{\rm{H}} \Gamma_0 \exp\left(-\tau\right)\,\df{\ell'}
    \label{eq:gw4}
\end{equation}
We next define a photon number-weighted average of the photoionisation cross section $\sigavg$ as
\begin{equation}
    \sigavg \equiv \frac{\int_{\nu_{\rm{th}}}^\infty\sigma(\nu)\phi(\nu)\,\df{\nu}}
        {\int_{\nu_{\rm{th}}}^\infty \phi(\nu)\,\df{\nu}}
    \label{eq:gw5}
\end{equation}
where $\phi(\nu) \equiv J(\nu)/h\nu$ gives the number of photons per second with frequency $\nu$.
With this definition, the total optical depth becomes
\begin{equation}
    \tau \equiv \int_0^\ell X_{\HI} \sigavg n_{\rm{H}}\,\df{\ell'} \quad \text{and so} \quad \df{\tau} = X_{\HI} \sigavg n_{\rm{H}}\,\df{\ell'}
    \label{eq:gw6}
\end{equation}
Inserting these results into Eq.~\ref{eq:gw4} gives
\begin{equation}
    \Sigma_{\Lya} = f_{\Lya} \frac{\Gamma_0}{\sigavg} \int_0^{\tau'=\tau} \exp(-\tau')\,\df{\tau'}
    \label{eq:gw7}
\end{equation}
In the optically thick limit, $\tau\rightarrow\infty$ and so $\Sigma_{\Lya} = f_{\Lya}\,\Gamma_0/\sigavg$. Rewriting this result using Eq.~\ref{eq:gw5} and the definition of $\Gamma_0$ previously given in Eq.~\ref{eq:gamma0} yields
\begin{equation}
    \Sigma_{\Lya} = 4 \pi f_{\Lya} \int_{\nu_{\rm{th}}}^\infty \phi(\nu)\,\df{\nu}
    \label{eq:gw_result}
\end{equation}
which is the \citet{gouldImagingForestLyman1996} result: in the optically thick limit, all photons are absorbed and therefore the $\Lya$ surface brightness tends to a constant fraction of the illuminating flux set by $f_{\Lya}$.
To obtain an equivalent result for $\Ha$, we multiply by the case B flux ratio $F_{\Lya} / F_{\Ha} = 8.5$ for $T = 10^4\,\text{K}$ \citep{osterbrockAstrophysicsGaseousNebulae2006}.
The resulting prediction, $\Sigma_{\Ha,\text{GW}} = 2.15 \times 10^{-20} \sbunit$, provides a useful check of our code.

\section{Additional tests of ionisation balance}
\label{app:moretests}

In this Appendix, we provide details of a number of additional comparisons with \textsc{cloudy} that were performed to verify the operation of the code described in Section~\ref{ssec:code}.

\begin{figure}
\subimport{figs/}{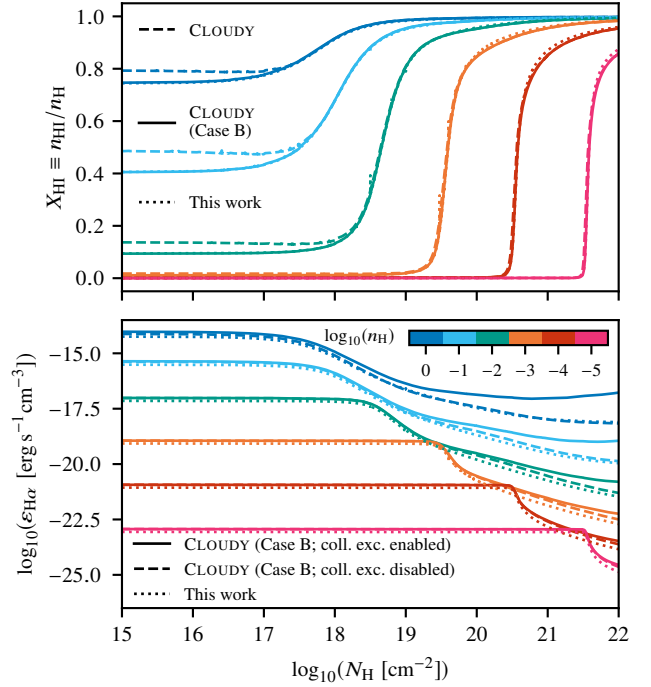}
\caption{\emph{Upper panel:} $\HI$ neutral fraction as a function of total H column density for isothermal, H-only plane parallel models. Results from our code are shown by dotted lines; over much of the plotted $N_{\text{H}}$ range, these are indistinguishable from the solid lines, which show \textsc{cloudy} results with the Case B assumption switched on. Relaxing this assumption results in the curves shown by dashed lines.
\emph{Lower panel:} $\Ha$ volume emissivity for the same models. Our models are again shown with dotted lines, while solid and dashed lines correspond to Case B \textsc{cloudy} models with collisional excitation processes enabled and disabled respectively.}
\label{fig:yprf_emis_noHe}
\end{figure}

Firstly, we consider the effects of our use of the case B recombination assumption, by comparing the hydrogen neutral fractions $X_{\HI} \equiv n_{\HI}/n_{\text{H}}$ that we obtain with corresponding predictions from \textsc{cloudy}, both using its default self-consistent treatment of recombination radiation, and overriding this behaviour to also assume the case B limit.
The upper panel panel of Fig.~\ref{fig:yprf_emis_noHe} indicates that this approximation results in a slight underestimate of the neutral fraction  at column densities $N_{\text{H}} < 10^{18}\,\text{cm}^{-2}$.

The volume emissivities plotted in the bottom panel of Fig.~\ref{fig:yprf_emis_noHe} appear to show that this optically thin gas contributes the majority of the $\Ha$ emission; however, this is an effect of the artificial constant-density assumption used in the slab models.
For the physically-motivated models we consider, the gas density profile is dictated by hydrostatic equilibrium so that $n_{H}$ rapidly decreases with increasing radius and the $\Ha$ emissivity peaks near the ionisation front.
This is located at $N_{\text{H}} \gtrsim 10^{18}\,\text{cm}^{-2}$, where the case B limit is applicable; accordingly, we do not expect the absence of self-consistent diffuse radiation in our code to significantly affect our results.
Fig.~\ref{fig:yprf_emis_noHe} also indicates emissivities below the \textsc{cloudy} values at high column and volume densities.
This is likely caused by the exclusion of collisional excitation processes in our code, as disabling these processes in \textsc{cloudy} produces the dashed curves in Fig.~\ref{fig:yprf_emis_noHe}, with which our results are a better match.
Since the emissivity of the predominantly neutral gas found at high $N_{\text{H}}$ is at least two orders of magnitude lower than that of gas near the ionisation front, we expect the net impact of this discrepancy on the total surface brightness to be minimal.

\begin{figure}
\subimport{figs/}{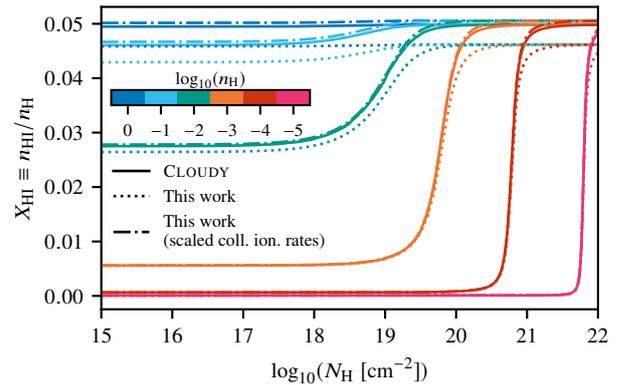}
\caption{$\HI$ neutral fraction as a function of total H column density for isothermal, H-only plane parallel models with $T_{\text{gas}}=20,000\,\text{K}$.
Solid and dotted curves correspond to output generated by \textsc{cloudy} and our code respectively, while the dot-dashed curves show our predictions if the collisional ionisation rates we use are rescaled to match those in \textsc{cloudy}.}
\label{fig:yprf_2e4k_noHe}
\end{figure}

Finally, in Fig.~\ref{fig:yprf_2e4k_noHe} we show results for a set of plane parallel models run at a higher gas temperature of $20,000\,\text{K}$.
At this temperature, collisional ionisation dominates throughout the slab, resulting in neutral fractions $\lesssim 0.05$.
There is a systematic offset of ${\sim} 10\%$ in the neutral fractions predicted by our code and by \textsc{cloudy}, which is due to a difference of similar magnitude in the collisional ionisation rate coefficients assumed by the two codes.
Rescaling the \textsc{chianti} rates we employ to match those reported by \textsc{cloudy} produces the dot-dashed curves in Fig.~\ref{fig:yprf_2e4k_noHe}, reducing the offset in the neutral fractions to below 1\%.
Hence, excluding differences in the input data used, we find excellent agreement between our code and \textsc{cloudy} in the collisionally-ionised limit.

In summary, while the simplicity of the code we present here as compared to \textsc{cloudy} results in some differences in the predicted values of $X_{\HI}$ and $\varepsilon_{\Ha}$, these occur for regimes that are either absent in our physically-motivated models, or contribute little to the observed surface brightness.
Therefore, our code's predictions appear to be robust in its intended domain of applicability.

\bsp	
\label{lastpage}
\end{document}